# The unexpected thermal conductivity from graphene disk, carbon nanocone to carbon nanotube

Dengke Ma[1,2,#], Hongru Ding[1,2,#], Han Meng[1,2], Xiaoman Wang[1,2], Nuo Yang[1,2,*], Xing Zhang[3,*]

[1]State Key Laboratory of Coal Combustion, Huazhong University of Science and Technology (HUST), Wuhan 430074, P. R. China

[2]Nano Interface Center for Energy (NICE), School of Energy and Power Engineering, Huazhong University of Science and Technology (HUST), Wuhan 430074, P. R. China

[3]Key Laboratory for Thermal Science and Power Engineering of Ministry of Education, Department of Engineering Mechanics, Tsinghua University, Beijing, P. R. China

[#]D. M. and H. D. contributed equally to this work.

*Corresponding authors. E-mail: nuo@hust.edu.cn (Nuo Yang) and
x-zhang@tsinghua.edu.cn (Xing Zhang)

**ABSTRACT**

Graphene and single-wall carbon nanotube (SWCNT) have attracted great attention because of their ultra-high thermal conductivity. However, there are few works exploring the relations of their thermal conductivity quantitatively. The carbon nanocone (CNC) is a graded structure fall in between graphene disk (GD) and SWCNT. We perform non-equilibrium molecular dynamics (NEMD) simulation to study the thermal conductivity of CNC with different apex angles, and then compare them with that of GD and SWCNT. Our results show that, different from the homogeneous thermal conductivity in SWCNT, the CNC also has a natural graded thermal conductivity which is similar to the GD. Unexpectedly, the graded rate keeps almost the same when the apex angle decreases from 180° (GD) to 19°, but then suddenly declines to zero when the apex angle decreases from 19° to 0° (SWCNT). What is more interesting, the graded effect is not diminished when the interatomic force constant is weakened and mean free path is shorten. That is, besides nanoscale, the graded effect can be observed in macroscale graphene or CNC structures.

## INTRODUCTION

Over the past decades, most of the researches have focused on graphene[1] and single-wall carbon nanotubes (SWCNT) [2] due to their unusually high thermal conductivity. It is found that the thermal conductivity of the suspended graphene exceeds (2500 + 1100/-1050) W/m-K near 350 K, [3-5] and the thermal conductivity of SWCNT exceeds 2000 W/m-K at room temperature. [6, 7] Although carbon nanocone (CNC) has been observed just after the SWCNT, [8, 9] it drew less attention. Researches towards CNC mainly concentrate on its growth mechanism, [9-13] mechanical properties [14-16], and electronic properties [17-19]. It is found that CNC can be used for atomic force microscopy, [20] as thermal rectifier, [21] ammonia sensor, [22] sorbent materials, [23] autonomous pump[24, 25] and interface materials. [26] However, few works have paid attention to its thermal properties [21] even though thermal properties are of fundamental and practical significance.

Different from previous studies on divergence of thermal conductivity in nano-structures with different size, [27-30] Yang et al found a novel graded thermal conductivity along the longitudinal direction in one graphene disk (GD) in 2015 based on molecular dynamics simulation.[31] And the similar phenomenon was observed by Wang et al through Monte Carlo simulation. [32] This new finding broadens our understanding of thermal transport property of materials in nanoscale, and opens up a new door for us to explore the connection of thermal conductivity between GD, CNC and SWCNT.

Graphene and SWCNT are strongly correlated with each other in structure. A SWCNT can be described as graphene that is rolled up into a seamless cylinder. Generally, the phonon dispersion relations in a SWCNT can be obtained from those of the 2D graphene sheet by using the zone folding approach. However, zone-folding of the graphene phonon branches does not always give the correct dispersion relation for a carbon nanotube.[33] CNC is characterized by the apex angle θ (as shown in Fig.1c). The extreme situations when the apex angle goes to 180°and 0°are corresponding to GD and SWCNT, respectively. So CNC has an outstanding structure advantage to relate GD and SWCNT together by changing its apex angle (as shown in Fig.1a)。

In this letter, the thermal conductivity of GD, CNC and SWCNT along the longitudinal direction were calculated by using the non-equilibrium molecular dynamics (NEMD) simulation. The dependence of thermal conductivity on apex angle was investigated. The results of CNCs were compared with those of GD and SWCNT. To understand the underlying physical mechanism, we analyzed the atomic position distribution function (APDF) and the phonon power spectra (PPS) of atoms close to the top and bottom atomic layers in GD, CNC and SWCNT.

## Structure and Methods

CNC is rolled up by cutting the GD a sector off (as shown in Fig. 1b).[9] When the apex angle decreases, its degree of asymmetry decreases (as shown in Fig. 1c). The lattice constant (a) and thickness (d) of GD, CNC and SWCNT are 0.1438 nm and 0.334 nm, respectively. The GD and CNC have 73 layers, and their interlayer spacing is 0.1438 nm (as shown in Fig. 1b). The SWCNT has 84 layers, and its interlayer spacing is 0.1245 nm.

When studying the thermal conductivity of CNC, we focus on the CNCs with the apex angle θ of 113°,84°,60°,39° and 19°, which are rolled up by cutting a GD 60°,120°,180°,240° and 300° off respectively (as shown in Fig. 1b). And these structures can be synthesized experimentally.[9, 10] As the thermal conductivity depends on system size[28, 30] and tube diameter[34], we keep the length L and the top radius $r_{top}$ (as shown in Fig. 1d) the same for different apex angle cases. Moreover, the same top radius makes it possible to realize all the structures by merely changing the apex angle from 180° to 0°. L is 10nm, and $r_{top}$ is approximately the same with the radius of a (10,10) SWCNT.

As we study the thermal conductivity of CNC by using the classical NEMD method, a temperature gradient is built in CNC from top to bottom along the longitudinal direction. The CNC is coupled with Nosé–Hoover thermostats at the 3rd to 7th and (N−6)th to (N−2)th layers with temperatures are $T_{top}$ and $T_{bottom}$, respectively. To ensure our results are independent of heat bath, Langevin thermostats are also used.[29] Because there is rectification effect for large temperature difference,[21] the normalized temperature difference ratio ($\Delta T/T_{average}$) is designed as 0.02 in all simulations, where the rectification effect can be ignored. And atoms at the boundaries (the 1st to 2nd and (N−1)th to Nth layers) are fixed. The optimized Tersoff potential is applied to describe the C-C interactions, which can better describe the lattice properties of GD and SWCNT.[35, 36] And our result of the thermal conductivity of a 10nm long (10, 10) SWCNT (as shown in Fig. S1) is 122.2 W/m-K, which is close to the values reported by others. [37, 38] To integrate the discretized differential equations of motions, the velocity Verlet algorithm is used. (MD simulation details are shown in S IV of the supporting information)

**Results and Discussion**

The temperature profiles of CNC along the longitudinal direction from top to bottom for different apex angles are plotted in Fig. 2. To avoid the non-linear effect from heat bath, we selected the atoms 1.87 nm far away from the heat bath to record their properties. The symbols are direct MD results. As is shown in Fig. 2a, the temperature profile of GD is a curve rather than a line. Clearly, the thermal conductivity of GD in our result is graded, similar to that of Yang et al.[31]

Same as Fig. 2a, the temperature profiles of CNC are also curves in Fig. 2b, 2c, 2d, 2e, 2f. It implies that, instead of being homogeneous, the thermal conductivities of CNCs with different apex angles are graded, too. It is because that the CNC forms a similar heat transfer to that in GD, streaming from inside to outside (as shown in Fig. 1d, the top view). The temperature profile of SWCNT is obviously a line (as shown in Fig. S1). Thus, the thermal conductivity of SWCNT is homogeneous, which matches well with previous work.[39]

Our previous analytical result showed that the non-homogeneous steady state temperature distributions and thermal conductivity of GDs satisfies: [31]

$$T(r) = \begin{cases} T(r_{in}) + [T(r_{out}) - T(r_{in})] \dfrac{R_{normal}(r)^{1-\alpha} - [\ln(r_{in}/C)/\ln(r_{out}/C)]^{1-\alpha}}{1 - [\ln(r_{in}/C)/\ln(r_{out}/C)]^{1-\alpha}}, & \alpha \neq 1 \\ \\ T(r_{in}) + [T(r_{out}) - T(r_{in})] \dfrac{R_{normal}(r) - [\ln(r_{in}/C)/\ln(r_{out}/C)]}{1 - [\ln(r_{in}/C)/\ln(r_{out}/C)]}, & \alpha = 1 \end{cases} \quad (1)$$

$$\kappa(r) = \kappa_0 \left[ \dfrac{\ln(C/r)}{\ln(C/r_{out})} \right]^\alpha = \kappa_0 [R_{normal}(r)]^\alpha$$

$$R_{normal}(r) = \dfrac{\ln(C/r)}{\ln(C/r_{out})} \quad (2)$$

where $r_{in}$ is the distance from the vertex of the cone to the top of the cone, $r_{out}$ is the distance from the vertex of the cone to the bottom of the cone (as shown in Fig. 1c). α is the exponential factor, which would depend on the apex angle, temperature and length. $\kappa_0$ and C are constants.

We use Eq. (1) to fit the temperature profile data obtained from MD, which may overcome the problem of fluctuation in temperature profile. As shown in Fig. 2, the numerical data can be well fitted by using Eq. (1). The adjusted R square for fitted lines are 0.999, 0.999, 1.000, 0.997, 0.998 and 0.996 corresponding to the apex angle of 180°, 113°, 84°, 60°, 39° and 19°, respectively. Using the fitted parameters, we get the thermal conductivity of GD and CNC by Eq. (S3) and Eq. (2). The calculation of the thermal conductivity of SWCNT is shown in SI of the supporting information.

As shown in Fig. 3 on a log-log plot, the thermal conductivity of SWCNT is homogeneous, while for GD and CNC, their thermal conductivity increase linearly from top to bottom along the longitudinal direction. Moreover, when decreasing the apex angle from 180° to 19°, the thermal conductivity increases obviously. Since we keep the top radius the same (as shown in Fig.1a), the CNC becomes thinner when the apex angle decreases. This tendency is similar to that in the SWCNT with different tube diameters.[40] Since CNC can be treated as SWCNT connected with different tube diameters, the thermal conductivity of CNC will increase when the apex angle decreases.

What is more interesting, the slopes of the thermal conductivity lines keep almost the same when the apex angle decreases from 180° (GD) to 19°, which indicates that the graded rate of the thermal conductivity keeps almost the same for different apex angle. Since α, the exponential factor in Eq. (2), indicates the graded rate of thermal conductivity, we focus on it to get a better understanding of this change. When α equals zero, it corresponds to a homogeneous thermal conductivity. As shown in Fig. 3b, the lowest α value presented by SWCNT is zero. While the graded rate α changes a little when the apex angle decreases from 180°(GD) to 19°, but then suddenly declines to zero when the apex angle decreases from 19° to 0° (SWCNT). This means the graded rate α is not sensitive to the changing of structural asymmetry. The structural transition from symmetrical SWCNT to asymmetrical CNC and GD is a qualitative change which corresponding to homogeneous and graded thermal conductivity.

To better comprehend the graded thermal conductivity and explore the reason of the unexpected change of α with the apex angle. We remove the heat bath, and keep the whole system in equilibrium state. Then we record the position and velocity of the atoms close to the top (16th) and bottom (58th) atomic layer of GD (inner and outside for GD), CNC and SWCNT (19th and 67th atomic layer for SWCNT), and calculate the APDF (as shown in Fig. 4, Fig. S2 and Fig. S3) and the PPS (as shown in Fig. S4). Here, the APDF and PPS are exhibited along three directions, two in-plane directions and one out-plane direction, as illustrated in Fig. 1c.

The APDFs along normal directions are calculated and shown in Fig. 4. The atoms close to the top atomic layer of CNC (inside of GD) are confined in small amplitude around their equilibrium positions. While, the atom close to the bottom (outside of GD) have a larger spread of vibration. The difference of amplitude comes from the difference of vibration modes existing on each atom. This is because the number of vibration modes is same for different atomic layers, which is observed at the PPSs (as shown in Fig. S4).[41, 42] While the atomic layers close to the top have a smaller number of atoms. So on average, there are more vibration modes existing on each atom close to the top. These vibrations modes destructively interact and result in a confined vibration and smaller amplitude for atoms close to the top. At the same time, the denser vibration modes take more interferences and scatterings, which leads to a smaller thermal conductivity for atoms close to the top. On the contrary, as shown in Fig. S2, the APDF of atoms close to the two ends in SWCNT are the same along three directions due to the same atom number. It agrees well with the homogeneous thermal conductivity of SWCNT.

With regard to the graded effect, we also studied how the phonon diffusive transport effects on the graded rate $\alpha$. The graphene has high thermal conductivities and long phonon mean free path (~775 nm).[43] Due to the limitation of simulation time, we can't simulate a GD whose radius is on the order of the phonon mean free path. However, we can reduce the phonon mean free path by weakening the interatomic force constant, because the weakening can increase the anharmonic effect and three-phonon scatterings which will reduce the mean free path. We calculated the corresponding graded rate $\alpha$ of GD and CNC with apex angle 60° corresponding to three different interatomic force (as shown in Fig. 4d). Interestingly, when the interatomic force constant is weakened, the graded rate keeps almost the same. There is obvious graded effect in the structure with short phonon mean free path, which declares that the super-diffusive transport is not responsible for the graded thermal conductivity. That is, besides nanoscale, the graded effect can be observed in macroscale graphene or CNC structures.

## Conclusion

We study the thermal conductivity of CNC with different apex angles, and then compare them with that of GD and SWCNT. Our NEMD simulation results indicate that the thermal conductivity of CNC along the longitudinal direction is graded, which is similar to GD, rather than homogeneous (SWCNT). We assert that the graded thermal conductivity comes from the difference of vibration modes existing on each atom. The denser vibration modes take more interferences and scatterings, which leads to a smaller thermal conductivity for atoms close to the top. What's more interesting, we find that the graded rate keeps almost the same when the interatomic force constant is weakened. That is, the graded effect is not dependent on ballistic transport and not limited in nanoscale. We find that the graded rate $\alpha$ does not depend significantly on the apex angle, but then suddenly declines to 0 (homogenous) when the structure changes to SWCNT. So the graded rate $\alpha$ is not sensitive to the changing of structural asymmetry. And the structural transition from symmetrical SWCNT to asymmetrical CNC and GD is a qualitative change which corresponding to homogeneous and graded thermal conductivity.

**Acknowledgments**

The project was supported by the National Natural Science Foundation of China 51327001 (XZ) and 51576067 (NY). The authors are grateful to Shiqian Hu, Xiaoxiang Yu, Yingying Zhang, Masato Ohnishi and Han Meng for useful discussions. The authors thank the National Supercomputing Center in Tianjin (NSCC-TJ) for providing help in computations.

Additional Information Supplementary information accompanies this paper.

**Competing financial interests:** The authors declare no competing financial interests.

# References

[1] K.S. Novoselov, A.K. Geim, S.V. Morozov, D. Jiang, Y. Zhang, S.V. Dubonos, I.V. Grigorieva, A.A. Firsov, Electric field effect in atomically thin carbon films, Science, 306 (2004) 666-669.
[2] S. Iijima, Helical microtubules of graphitic carbon, Nature, 354(6348) (1991) 56-58.
[3] W. Cai, A.L. Moore, Y. Zhu, X. Li, S. Chen, L. Shi, R.S. Ruoff, Thermal transport in suspended and supported monolayer graphene grown by chemical vapor deposition., Nano Lett., 10 (2010) 1645-1651.
[4] A.A. Balandin, Thermal properties of graphene and nanostructured carbon materials, Nat. Mater., 10 (2011) 569-581.
[5] H. Wang, K. Kurata, T. Fukunaga, H. Takamatsu, X. Zhang, T. Ikuta, K. Takahashi, T. Nishiyama, H. Ago, Y. Takata, In-situ measurement of the heat transport in defect- engineered free-standing single-layer graphene, Scientific Reports, 6 (2016) 21823.
[6] M. Fujii, X. Zhang, H. Xie, H. Ago, K. Takahashi, T. Ikuta, H. Abe, T. Shimizu, Measuring the thermal conductivity of a single carbon nanotube, Phys. Rev. Lett., 95(6) (2005) 065502.
[7] P. Kim, L. Shi, A. Majumdar, P.L. McEuen, Thermal Transport Measurements of Individual Multiwalled Nanotubes, Phys. Rev. Lett., 87(21) (2001) 215502.
[8] S. Iijima, T. Ichihashi, Y. Ando, Pentagons, heptagons and negative curvature in graphite microtubule growth, Nature, 356(6372) (1992) 776-778.
[9] M. Ge, K. Sattler, Observation of fullerene cones, Chemical physics letters, 220(3) (1994) 192-196.
[10] A. Krishnan, E. Dujardin, M.M.J. Treacy, J. Hugdahl, S. Lynum, T.W. Ebbesen, Graphitic cones and the nucleation of curved carbon surfaces, Nature, 388(6641) (1997) 451-454.
[11] V.I. Merkulov, M.A. Guillorn, D.H. Lowndes, M.L. Simpson, E. Voelkl, Shaping carbon nanostructures by controlling the synthesis process, Appl. Phys. Lett., 79(8) (2001) 1178-1180.
[12] C.-T. Lin, C.-Y. Lee, H.-T. Chiu, T.-S. Chin, Graphene structure in carbon nanocones and nanodiscs, Langmuir, 23(26) (2007) 12806-12810.
[13] Z. Tsakadze, I. Levchenko, K. Ostrikov, S. Xu, Plasma-assisted self-organized growth of uniform carbon nanocone arrays, Carbon, 45(10) (2007) 2022-2030.
[14] S.P. Jordan, V.H. Crespi, Theory of carbon nanocones: mechanical chiral inversion of a micron-scale three-dimensional object, Phys. Rev. Lett., 93(25) (2004) 255504.
[15] J. Wei, K.M. Liew, X. He, Mechanical properties of carbon nanocones, Appl. Phys. Lett., 91(26) (2007) 261906.
[16] Y.-G. Hu, K.M. Liew, X. He, Z. Li, J. Han, Free transverse vibration of single-walled carbon nanocones, Carbon, 50(12) (2012) 4418-4423.
[17] J.-C. Charlier, G.-M. Rignanese, Electronic Structure of Carbon Nanocones, Phys. Rev. Lett., 86(26) (2001) 5970-5973.
[18] O. Shenderova, B. Lawson, D. Areshkin, D. Brenner, Predicted structure and electronic properties of individual carbon nanocones and nanostructures assembled from nanocones, Nanotechnology, 12(3) (2001) 191.
[19] Q. Wang, R. Gao, S. Qu, J. Li, C. Gu, Metallic electrical transport in inter-graphitic planes of an individual tubular carbon nanocone, Nanotechnology, 20(14) (2009) 145201.
[20] I.-C. Chen, L.-H. Chen, X.-R. Ye, C. Daraio, S. Jin, C.A. Orme, A. Quist, R. Lal, Extremely sharp carbon nanocone probes for atomic force microscopy imaging, Appl. Phys. Lett., 88(15) (2006) 153102.
[21] N. Yang, G. Zhang, B. Li, Carbon nanocone: A promising thermal rectifier, Appl. Phys. Lett., 93(24) (2008) 243111.
[22] M.T. Baei, A.A. Peyghan, Z. Bagheri, Carbon nanocone as an ammonia sensor: DFT studies, Structural Chemistry, 24(4) (2013) 1099-1103.
[23] J.M. Jiménez-Soto, S. Cárdenas, M. Valcárcel, Evaluation of carbon nanocones/disks as sorbent material for solid-phase extraction, Journal of Chromatography A, 1216(30) (2009) 5626-5633.
[24] Z.-c. Xu, D.-q. Zheng, B.-q. Ai, W.-r. Zhong, Autonomous pump against concentration gradient, Scientific Reports, 6 (2016) 23414.
[25] K. Xiao, G. Xie, Z. Zhang, X.Y. Kong, Q. Liu, P. Li, L. Wen, L. Jiang, Enhanced Stability and Controllability of an Ionic Diode Based on Funnel‐Shaped Nanochannels with an Extended Critical Region, Adv. Mat.,　(2016).
[26] M. Knaapila, O.T. Rømoen, E. Svåsand, J.P. Pinheiro, Ø.G. Martinsen, M. Buchanan, A.T. Skjeltorp, G. Helgesen, Conductivity enhancement in carbon nanocone adhesive by electric field induced formation of aligned assemblies, ACS applied materials & interfaces, 3(2) (2011) 378-384.
[27] B. Li, J. Wang, Anomalous Heat Conduction and Anomalous Diffusion in One-Dimensional Systems, Phys. Rev. Lett., 91(4) (2003) 044301.
[28] C.-W. Chang, D. Okawa, H. Garcia, A. Majumdar, A. Zettl, Breakdown of Fourier's law in

nanotube thermal conductors, Phys. Rev. Lett., 101(7) (2008) 075903.
[29] N. Yang, G. Zhang, B. Li, Violation of Fourier's law and anomalous heat diffusion in silicon nanowires, Nano Today, 5(2) (2010) 85-90.
[30] X. Xu, L.F.C. Pereira, Y. Wang, J. Wu, K. Zhang, X. Zhao, S. Bae, C. Tinh Bui, R. Xie, J.T.L. Thong, B.H. Hong, K.P. Loh, D. Donadio, B. Li, B. Özyilmaz, Length-dependent thermal conductivity in suspended single-layer graphene, Nature Communications, 5 (2014) 3689.
[31] N. Yang, S. Hu, D. Ma, T. Lu, B. Li, Nanoscale Graphene Disk: A Natural Functionally Graded Material–How is Fourier's Law Violated along Radius Direction of 2D Disk, Scientific Reports, 5 (2015) 14878.
[32] Z. Wang, R. Zhao, Y. Chen, Monte Carlo simulation of phonon transport in variable cross-section nanowires, Sci. China Ser. E, 53(2) (2010) 429-434.
[33] M.S. Dresselhaus, P.C. Eklund, Phonons in carbon nanotubes, Adv. Phys., 49(6) (2000) 705-814.
[34] J.A. Thomas, R.M. Iutzi, A.J.H. McGaughey, Thermal conductivity and phonon transport in empty and water-filled carbon nanotubes, Phys. Rev. B, 81(4) (2010) 045413.
[35] L. Lindsay, D.A. Broido, Optimized Tersoff and Brenner empirical potential parameters for lattice dynamics and phonon thermal transport in carbon nanotubes and graphene, Phys. Rev. B, 81(20) (2010) 205441.
[36] D. Ma, H. Ding, H. Meng, L. Feng, Y. Wu, J. Shiomi, N. Yang, Nano-cross-junction effect on phonon transport in silicon nanowire cages, Phys. Rev. B, 94(16) (2016) 165434.
[37] Q. Liao, Z. Liu, W. Liu, C. Deng, N. Yang, Extremely High Thermal Conductivity of Aligned Carbon Nanotube-Polyethylene Composites, Scientific Reports, 5 (2015) 16543.
[38] J.R. Lukes, H. Zhong, Thermal conductivity of individual single-wall carbon nanotubes, J. Heat Transf., 129(6) (2007) 705-716.
[39] S. Maruyama, A molecular dynamics simulation of heat conduction in finite length SWNTs, Physica B, 323(1–4) (2002) 193-195.
[40] S.-Y. Yue, T. Ouyang, M. Hu, Diameter Dependence of Lattice Thermal Conductivity of Single-Walled Carbon Nanotubes: Study from Ab Initio, Scientific Reports, 5 (2015) 15440.
[41] J. Zhang, Y. Hong, Y. Yue, Thermal transport across graphene and single layer hexagonal boron nitride, J. Appl. Phys., 117(13) (2015) 134307.
[42] V.P. Sokhan, D. Nicholson, N. Quirke, Phonon spectra in model carbon nanotubes, J. Chem. Phys., 113(5) (2000) 2007-2015.
[43] S. Ghosh, I. Calizo, D. Teweldebrhan, E.P. Pokatilov, D.L. Nika, A.A. Balandin, W. Bao, F. Miao, C.N. Lau, Extremely high thermal conductivity of graphene: Prospects for thermal management applications in nanoelectronic circuits, Appl. Phys. Lett., 92 (2008) 151911.

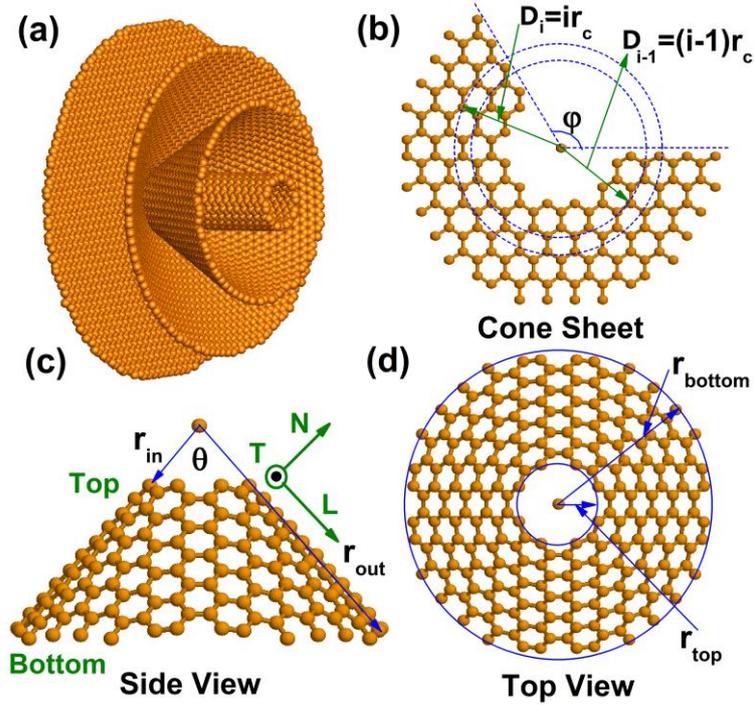

Figure 1. (Color online) (a) Schematic picture of the relationship of graphene disk, CNC and SWCNT. (b) A graphene disk which is cut a sector off, $\varphi$ is the angle of the sector cut off. $D$ is the distance from the atom to the center point in the cone sheet. We define the atoms in the $i$th layer as the atoms whose $D \subset [(i-1)r_c, ir_c]$. $r_c$ is the equilibrium carbon-carbon bond length. (c) The side view of CNC. $r_{in}$ and $r_{out}$ are the distance from the vertex of the cone to the top and bottom of the cone along the longitudinal direction. The length along the side face of nanocone is $L = (r_{out} - r_{in}) = Nr_c$, where N is the number of layers. $\theta$ is the angle of the cone. It also shows the transversal (T), normal (N) and longitudinal (L) directions. (d) The top view of CNC. $r_{top}$ and $r_{bottom}$ are the radius of the top and bottom circle of the cone respectively.

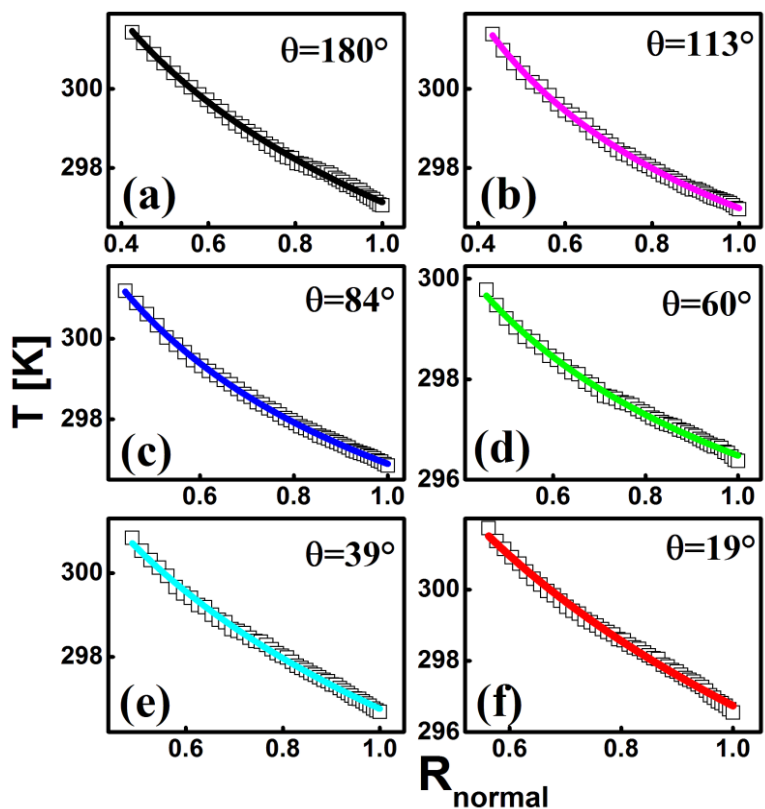

Figure 2. The temperature profiles along the longitudinal direction of graphene disk (a) and CNC with different apex angle as 113 °(b), 84 °(c), 60 °(d), 39 °(e) and 19 °(f) at 300K. The symbols are numerical data obtained by MD simulation, and the lines are fitted lines based on Eq.(1). The normalized radius, $R_{normal}$, is defined in Eq.(2). The temperature profile of SWCNT is shown in Fig. S1 (supplementary information).

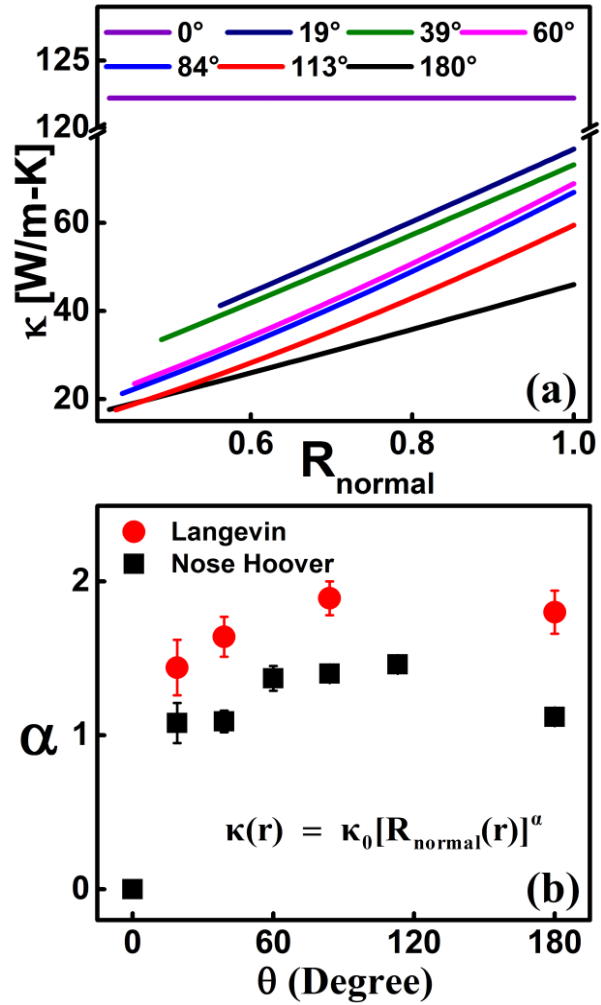

Figure 3. (a) The thermal conductivity of graphene disk (corresponding to 180°), CNCs with different apex angle and SWCNT (0°) at 300K. (b) The exponential factor α (graded rate) versus apex angle by Nosé–Hoover and Langevin heat bathes. The values by the two kinds of heat bath are consistent with each other, which show that the results are independent of heat bath.

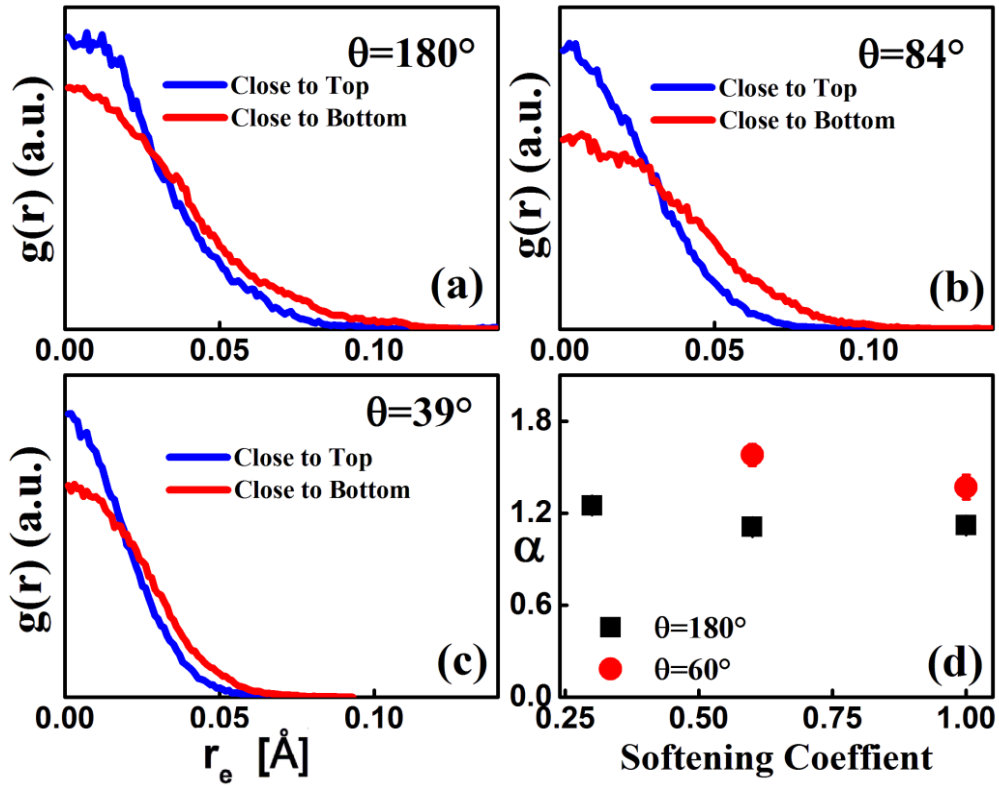

Figure 4. The APDF around its equilibrium positions along normal directions between atoms close to the top (inner for GD) and bottom (out for GD) atomic layers for different apex angles in equilibrium state, where $r_e$ is the distance to the atom equilibrium position. The apex angle is 180° (corresponding to GD), 84°, and 39° respectively. The rest cases along the longitudinal direction and the transverse direction are shown in Fig.S3. And the APDF of SWCNT is shown in Fig. S2 (supplementary information).

# The unexpected thermal conductivity from graphene disk, carbon nanocone to carbon nanotube

Dengke Ma[1,2,#], Hongru Ding[1,2,#], Han Meng[1,2], Xiaoman Wang[1,2], Nuo Yang[1,2,*], Xing Zhang[3,*]

[1]State Key Laboratory of Coal Combustion, Huazhong University of Science and Technology (HUST), Wuhan 430074, P. R. China

[2]Nano Interface Center for Energy (NICE), School of Energy and Power Engineering, Huazhong University of Science and Technology (HUST), Wuhan 430074, P. R. China

[3]Key Laboratory for Thermal Science and Power Engineering of Ministry of Education, Department of Engineering Mechanics, Tsinghua University, Beijing, P. R. China

*Corresponding authors: N.Y.(nuo@hust.edu.cn) and X.Z.(x-zhang@tsinghua.edu.cn)

[#]D. M. and H. D. contributed equally to this work.

**SI. The temperature profile and the atomic position distribution function (APDF) of (10,10) single-wall carbon nanotubes (SWCNT).**

As the temperature profile of SWCNT is line obviously, we use linear fit to get the slope of the line, and calculate the thermal conductivity of the (10,10) SWCNT according to Eq. (S1).

$$\kappa = -\frac{J}{C \cdot d \cdot \nabla T} \qquad (S1)$$

where x is distance along the axis direction of the tube. C is the perimeter of the tube. And d is the thickness.

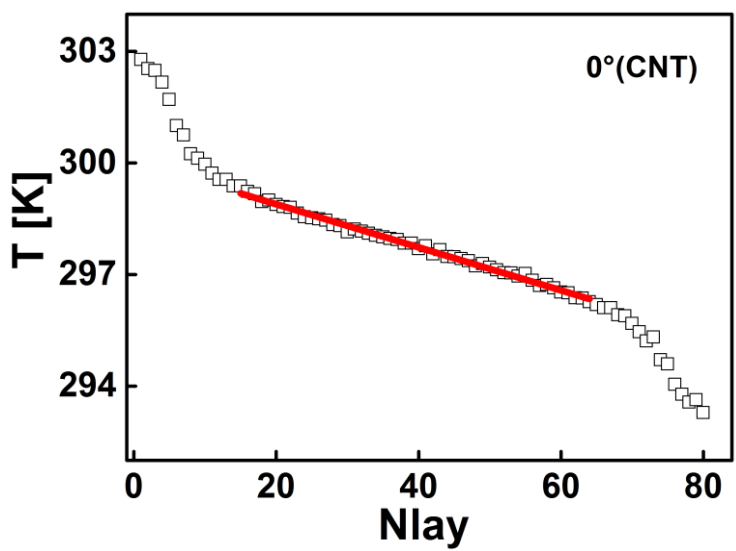

Figure S1. The temperature profiles of a (10,10) SWCNT. The symbols are MD results directly. The red line is fitted line.

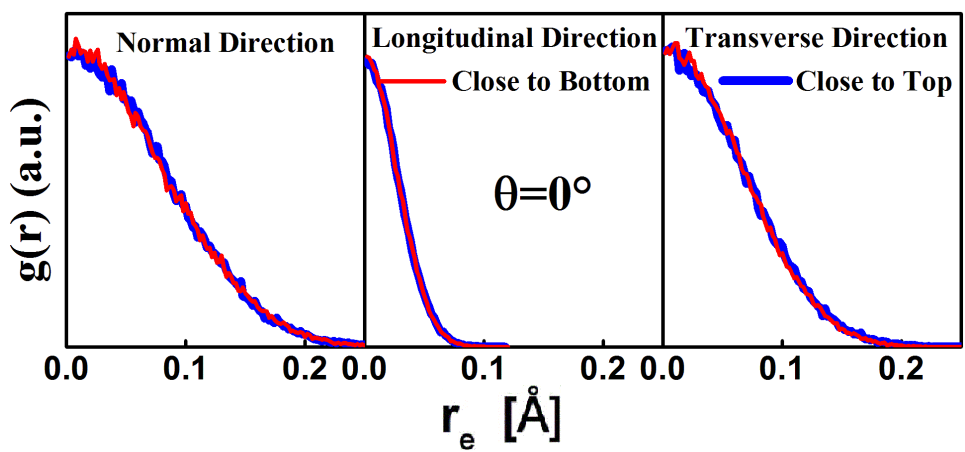

Figure S2 The APDF around its equilibrium positions along three directions between atoms close to the top and bottom atomic layers for SWCNT in equilibrium state, where $r_e$ is the distance to the atom equilibrium position.

**SII. The APDF for GD and CNC in equilibrium state**.

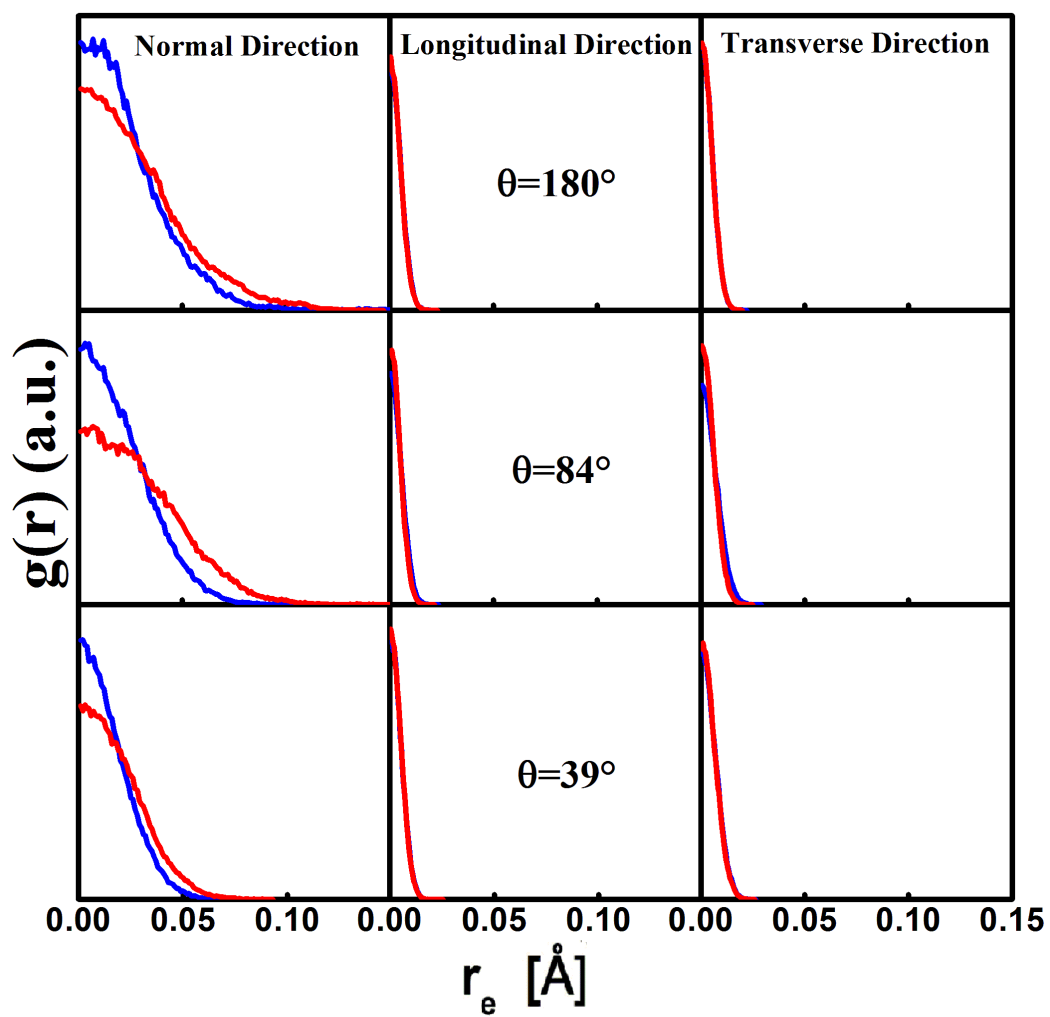

Figure S3 The APDF around its equilibrium positions along three directions between atoms close to the top and bottom atomic layers for different apex angle in equilibrium state, where $r_e$ is the distance to the atom equilibrium position. The apex angle is 180°(GD), 84°, and 39° respectively.

**SIII. The phonon power spectra for GD, CNC and SWCNT in equilibrium state**.

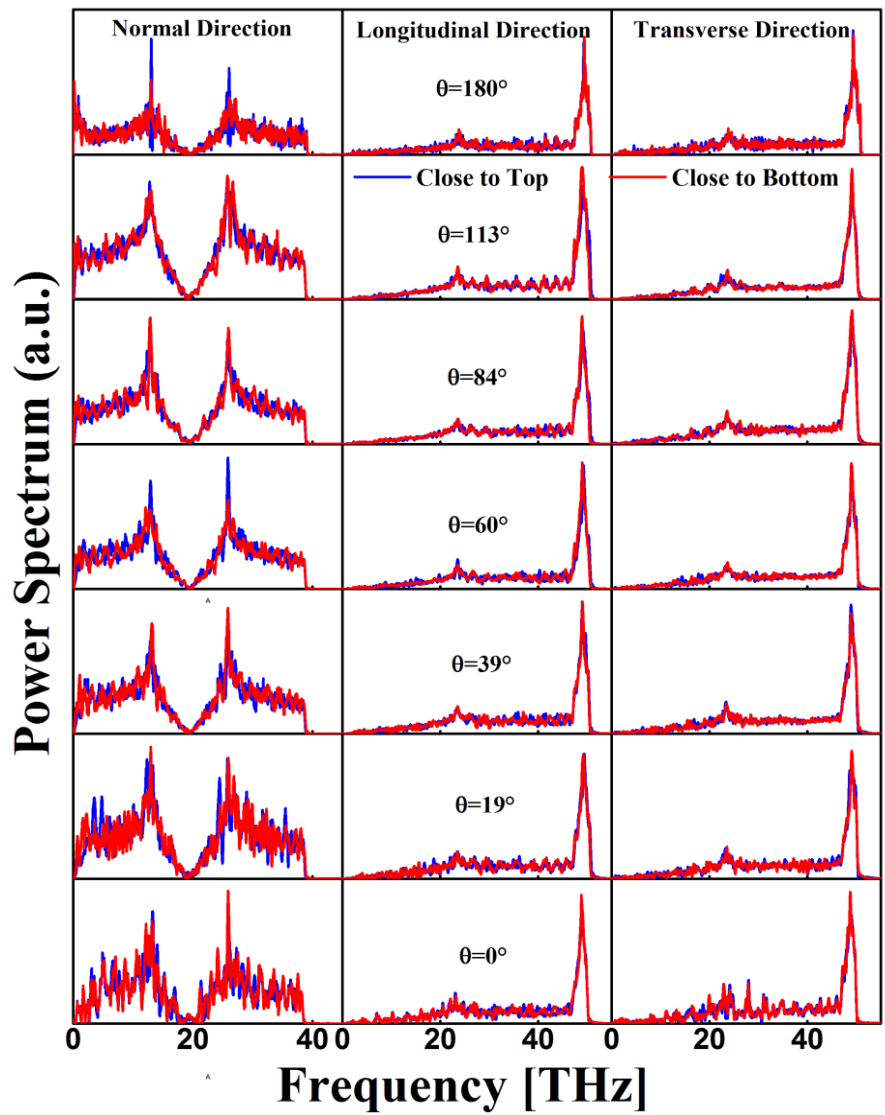

Figure S4. The phonon power spectra along three directions between atoms close to the top and bottom atomic layers according to Figure 1c in equilibrium state. The apex angle is 180 °(GD), 113 °, 84 °, 60 °, 39 °, 19 °and 0 °(SWCNT) respectively.

## SIV. The details of molecular dynamics (NEMD) simulation.

Initially, the system is equilibrated under the isothermal–isobaric ensemble (NPT) at the target temperature and 0 bar for 1 ns, followed by relaxation under a microcanonical ensemble (NVE) for 0.5 ns. Then temperature of hot and cold heat baths are set as 306 K and 296 K, respectively. The simulations are performed long enough to allow the system to reach a non-equilibrium steady state where the heat current going through the system is time independent. All results given here are obtained by averaging about 2 x $10^7$ time steps, and each time step is set as 0.5 fs. In general, the temperature $T_{MD}$ is calculated from the kinetic energy of atoms ($3Nk_B T_{MD}/2 = \sum_i m_i v_i^2/2$). The heat flux J along the nanocone can be calculated by averaging the added/removed energy at the two heat bath regions as:

$$J_{T_{out}(T_{in})} = \frac{1}{N_{T_{out}(T_{in})}} \sum_{i=1}^{N_{T_{out}(T_{in})}} \frac{\Delta \varepsilon_i}{2\Delta t} \tag{S2}$$

where $\Delta\varepsilon$ is the energy added to/removed from each heat bath ($T_{in}$ or $T_{out}$) at each step $\Delta t$. The thermal conductivity ($\kappa$) is calculated based on the Fourier definition as:

$$J = -2\pi \kappa \mathrm{d}r \sin(\theta/2) \frac{dT}{dr} \tag{S3}$$

where r is the distance to the vertex of the cone, and d is the thickness of GD, CNC and SWCNT. We use a combination of time and ensemble sampling to obtain better average statistics. The results represented are averaged from 6 independent simulations with different initial conditions.